\documentclass{article} 
\usepackage{iclr2020_conference,times}

\usepackage{amsmath,amsfonts,bm}

\def\eqref#1{equation~\ref{#1}}

\def\1{\bm{1}}

\DeclareMathAlphabet{\mathsfit}{\encodingdefault}{\sfdefault}{m}{sl}
\SetMathAlphabet{\mathsfit}{bold}{\encodingdefault}{\sfdefault}{bx}{n}

\usepackage{adjustbox}
\usepackage{booktabs}
\usepackage{tabu}

\usepackage{hyperref}
\hypersetup{
    colorlinks=false,
    linkcolor=blue,
    filecolor=magenta,      
    urlcolor=cyan,
}
\usepackage{url}

\title{A study of deep learning colon cancer \\detection in limited data access scenarios}

\author{Apostolia Tsirikoglou$^1$, Karin Stacke$^{1,3}$, Gabriel Eilertsen$^{1,2}$, Martin Lindvall$^{1,2,3}$, \\ \textbf{Jonas Unger}$^{1,2}$ \\ \\
$^1$~Department of Science and Technology, Link\"oping University, Sweden \\
\texttt{\{apostolia.tsirikoglou,karin.stacke,gabriel.eilertsen,} \\
\texttt{ martin.lindvall,jonas.unger\}@liu.se} \\
$^2$~Center for Medical Image Science and Visualization, Link\"oping University, Sweden \\
$^3$~Sectra AB, Sweden  \\
}

\iclrfinalcopy 
\begin{document}

\maketitle
\vspace{-.5cm}
Digitization of histopathology slides has led to several advances, from easy data sharing and collaborations to the development of digital diagnostic tools. Deep learning methods for classification and detection have shown great potential, but often require large amounts of training data that are hard to collect, and annotate. For many cancer types, the scarceness of data creates barriers for training deep learning models. One such scenario relates to detecting tumor metastasis in lymph node tissue, where the low ratio of tumor to non-tumor cells makes the diagnostic task hard and time-consuming. Deep learning-based tools can allow faster diagnosis, with potentially increased quality. Unfortunately, due to the sparsity of tumor cells, annotating this type of data demands a high level of effort from pathologists. Using weak annotations from slide-level images have shown great potential~\citep{campanella_clinical-grade_2019}, but demand access to a substantial amount of data as well.

In this study, we investigate mitigation strategies for limited data access scenarios. Particularly, we address whether it is possible to exploit mutual structure between tissues to develop general techniques, wherein data from one type of cancer in a particular tissue could have diagnostic value for other cancers in other tissues. Our case is exemplified by a deep learning model for metastatic colon cancer detection in lymph nodes. Could such a model be trained with little or even no lymph node data? As alternative data sources, we investigate 1) tumor cells taken from the primary colon tumor tissue, and 2) cancer data from a different organ (breast), either as is, or transformed to the target domain (colon) using Cycle-GANs~\citep{Zhu_2017_ICCV}. We show that the suggested approaches make it possible to detect cancer metastasis with no or very little lymph node data, opening up for the possibility that existing, annotated histopathology data could generalize to other domains. 

\section*{Data, Experiments and Results}
To exemplify our approach, we utilize data from two globally common types of cancer, breast and colon carcinoma, both originating from epithelial cells. This can make them visually similar despite originating from different organs. Specifically, we utilized the widely used, and publicly available CAMELYON17 breast cancer dataset~\citep{camelyon17}, and the recently released LNCO colon cancer dataset~(\cite{maras_regional_2019}). CAMELYON17 consists of sentinel lymph node images from five different medical centers in The Netherlands. From this dataset we used the 50 exhaustively annotated whole-slide images, corresponding to 43 patients. The LNCO dataset contains images from both primary and lymph node tumor samples from two medical centers in Sweden, and is annotated in a non-exhaustive manner. From LNCO we chose 37 patients. The train/test split was done at patient-level, with 32/5 patient cases respectively. The data from both datasets was extracted at a resolution of 0.5 microns per pixel, with patch size $256\times256$. All datasets were class-balanced to have equal amount of samples from each class. Henceforth, the notion \textit{non-tumor} and \textit{tumor} denotes normal and tumor data from colon lymph node tissue respectively, since the target task is to detect metastatic colon cancer in lymph nodes.

In the context of building data strategies for deep learning aided diagnostic tools when very limited amount of target training data is accessible, we define experiments based on three annotation effort categories: 1) \textit{high annotation effort}: lymph nodes from all patients, 2) \textit{low annotation effort}: primary tumor cells from all patients, and 3) \textit{medium annotation effort}: primary tumor cells from all patients, along with a smaller subset of lymph node samples, coming from a very limited number of patients. To validate our study we divided the LNCO train set in four disjoint groups of eight patients each, which were used in all experiments. Each group has approximately the same amount of primary tumor patches, meaning that in the low effort scenario $25\%$ of LNCO primary tumor train set is used. In the high effort experiment we balanced the amount of lymph node patches to be similar for all of the four groups, and over all patients. Finally, in the medium effort scenario each group utilizes all the available primary tumor patches, while the lymph node patches came from only two patients. The tumor classifier was a simple convolutional neural network with three convolutional layers and one subsequent fully connected layer. Each one of the conducted experiments was trained five times to ensure proper statistical variation. 

\begin{figure}[ht!]
    \begin{center}
        \includegraphics[width=0.8\textwidth]{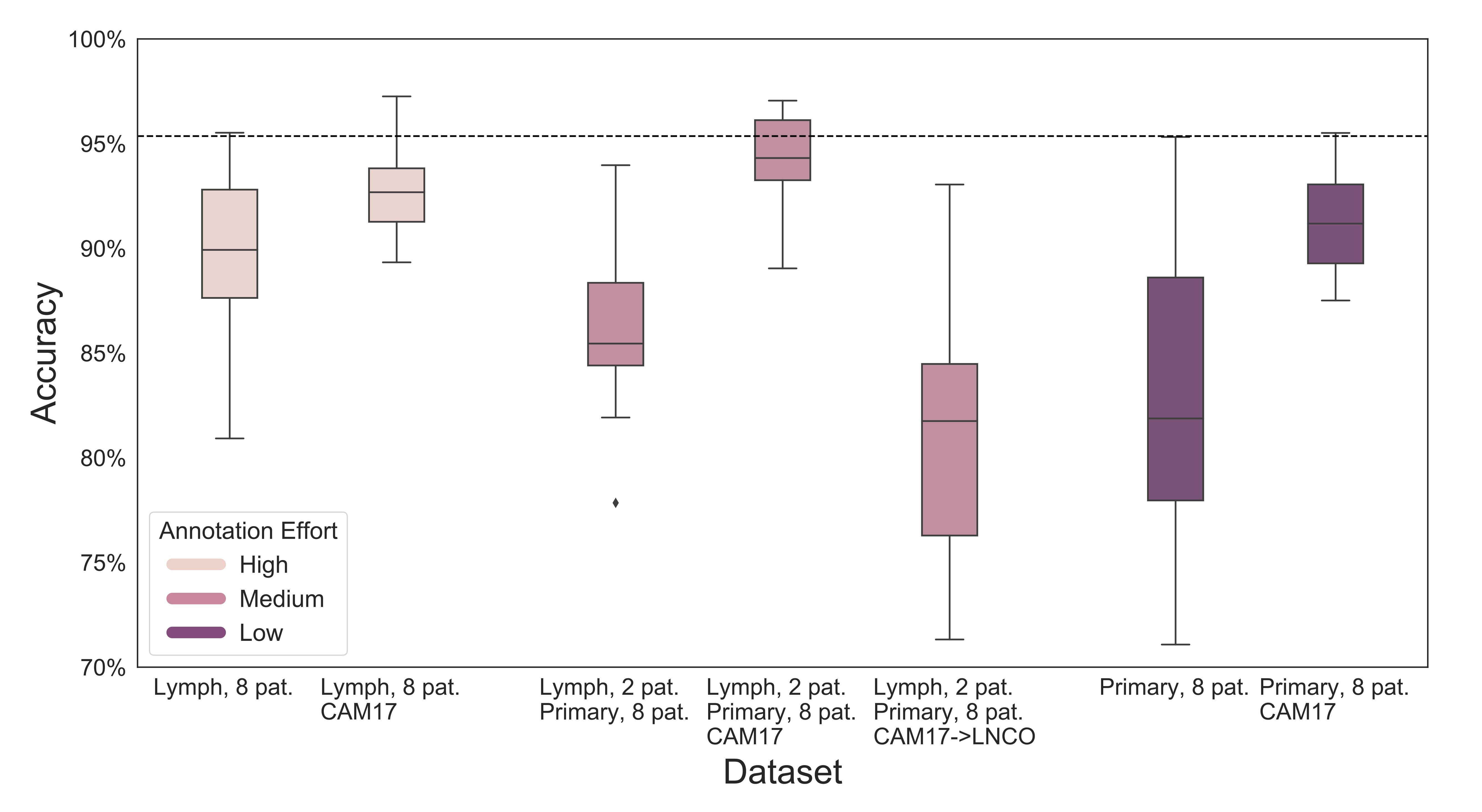}
    \end{center}
    \vspace{-.5cm}
    \caption{Patch-level accuracy for  all three annotation effort experiments (color coded, decresing effort from left to right), and training data augmentation strategies in LNCO subsets. Augmenting the LNCO subsets with CAMELYON17 always leads to increased performance, and in particular for medium annotation effort even exceeds the whole LNCO dataset baseline (dashed line). Note that "pat." stands for patients, "CAM17" for CAMELYON17, and "CAM17-$>$LNCO" for the Cycle-GAN domain adaption from breast to colon domain. }
    \label{fig:results}
\end{figure}

It is clear from Figure~\ref{fig:results} that using a subset of the LNCO train set, for any of the three annotation effort scenarios (color coded in the figure), is not as efficient as using the whole, diverse dataset (dashed line in the figure). Driven by recent studies of utilizing one organ's histopathology images to detect tumor cells in another~(\cite{Khan_prostate_breast_2019}), we augment the target dataset with an additional data source, which was selected as randomly and class-balanced patches from CAMELYON17. In this case the annotation effort remained the same. This simple strategy outperformed the total LNCO baseline for the medium annotation effort scenario, opening up possibilities in cross-domain cancer detection when limited data are available. As an alternative augmentation strategy, we transformed the CAMELYON17 patches to the colon domain using Cycle-GANs~(\cite{Zhu_2017_ICCV}), with the hypothesis that the transformed data would be closer to the target domain and thus boost performance. However, as it is shown in Figure~\ref{fig:results} this was not the case. We reason that this behaviour is due to the limited amount of data used when training the Cycle-GAN, but further investigation is needed. Please note that Figure~\ref{fig:results} reports patch-level accuracy to provide a relative measure for the expected performance between the experiment categories and training set strategies.  
\vspace{-.3cm}

\vspace{-.5cm}
\section*{Discussion}
We investigated training data strategies for metastatic colon cancer detection in limited data access scenarios. Limited training data access may occur due to either poor data collection resources, small data availability for a specific cancer type, or strenuous annotation processes.

The outcome of this work is twofold: 1) it is promising that adding primary tumor data in a small lymph node set with very limited number of patients leads to improved performance, and 2) it is possible to transfer histopathology information from one cancer type to another. The first finding enables primary tumor, that is often available with the lymph node section, to be considered as an additional training data source, while the second one opens up possibilities in cross-domain cancer detection. 

This study does not aim to present state-of-the-art performance on colon lymph node tumor detection. It is possible that transfer learning, more sophisticated networks or other types of augmentations yield better results. However, the study aims to explore different paths which might have been overlooked before, with the hope that this research will inspire others to continue along this line. There is great potential in leveraging already available data, to further increase diversity of available deep learning tools for clinical diagnostics, and thus lessen the disparity of fast and high quality cancer care across different cancer types worldwide.

\bibliography{iclr2020_conference}
\bibliographystyle{iclr2020_conference}

\end{document}